# Computational Assessment of Turbulent Eddy Impact on Hydrodynamic Mixing in a Stirred Tank Bioreactor with Vent-based Impellers

Ayodele James Oyejide [1*], Chidera Samuella Okeke [1], Jesuloluwa Emmanuel Zaccheus [1], Ebenezer Olubunmi Ige [2*]

[1] *Department of Biomedical Engineering, Afe Babalola University, Ado-Ekiti 360231, Nigeria*

[2] *Department of Mechanical Engineering, Rochester Institute of Technology, NY 14623, USA*

*Corresponding author: ebenezer.ige@rit.edu

**Abstract**

Homogeneity and efficient oxygen transfer are crucial for aerobic cultures, which is popularly performed in Stirred Tank Bioreactors, through internal mechanical agitation of the impellers. Although there are a number of impeller designs for achieving this purpose, there are still concerns about the ability of the impellers to yield homogeneity and mitigate or eliminate stagnant zones. In this study, a novel impeller design, with auxiliary agitators in form of vents, was introduced and evaluated for small lab-scale bioreactors. For the evaluation, 3D models of a single and double impeller configurations, placed in two different bioreactors were developed. Computational fluid dynamics was employed to carry out the hydrodynamic simulation using k-$\epsilon$ standard model in the bioreactors. Computational variables such as the flow velocity, streamlines, pressure and wall shear stress (on the shaft and impellers), eddy viscosity, turbulence eddy dissipation and turbulence kinetic energy were obtained and compared in both bioreactors to evaluate the performances at speeds of 50, 100 and 150 RPM. A comparison of the results with traditional segment-segment and segment-Rushton impellers shows that our double impeller configuration performs more desirably at speeds ranging from 100 to 150 RPM. Homogeneity was also achieved in both bioreactors, and there was significant reduction of stagnant zone (≤ 99%) in the double impeller configuration and significant mitigation in the single impeller agitation.

1. Introduction

Stirred bioreactors play a vital role in various biotechnological processes, including cell culture, fermentation, and production of bio-based products (Jia *et al* 2017). The performance of these bioreactors is largely influenced by the hydrodynamic behavior within the vessel which affects mixing efficiency, mass transfer rates, and shear stress distribution (Nadal-Rey *et al* 2022). Bioreactors generally work on the principle of fluid agitation, which is propelled by impellers; the traditional ones being the Rushton turbines, pitched blade turbines, and marine propellers (Jaszczur and Młynarczykowska 2020).

The Rushton turbine is one of the most commonly studied impeller designs in hydrodynamic studies. It consists of several flat blades attached to a central shaft. Hydrodynamic studies have shown that Rushton turbines induce radial flow patterns and generate significant turbulence in the bioreactor, resulting in efficient mixing and mass transfer (Ameur 2018). However, excessive shear stress levels associated with Rushton turbines can be detrimental to shear-sensitive cells or fragile biomolecules (Gelves 2020). The pitched blade turbine (PBT) is another widely investigated impeller design. PBTs feature blades set at an angle to the impeller shaft and studies have demonstrated that PBTs promote axial flow patterns and generate lower shear stress compared to Rushton turbines (Mollinedo 2022, Fan and Luan 2012, Li and Wang 2022). This impeller design is often preferred for shear-sensitive applications, such as mammalian cell culture, where maintaining cell viability and product integrity is critical. Furthermore, Marine propellers, commonly used in large-scale bioreactors, have been the subject of hydrodynamic studies to assess their performance characteristics. These impellers typically consist of several blades shaped like propeller blades. Studies have shown that marine propellers induce strong axial flow patterns and provide efficient mixing and gas-liquid mass transfer (Jaszczur and Młynarczykowska 2020). Their design enables effective circulation of the culture medium and homogeneous distribution of nutrients and gases within the bioreactor. One key limitation is their high shear stress generation. In addition, Marine propellers are known for their strong axial flow and high shear rates, which can be detrimental to sensitive cell cultures and delicate biomolecules (Brumley *et al* 2015). They also create stagnant zones (Vardhan *et al* 2019).

Optimizing the impeller design is crucial for achieving desired process outcomes, since design is an important aspect of product optimization (Singh *et al* 2021, Ayodele *et al* 2021, Adetola and Oyejide 2015). While existing impeller designs have made significant contributions to bioreactor performance, they are not without limitations. Challenges such as inadequate mixing, high shear stress, scalability issues, oxygen transfer limitations, and foaming persist in certain applications. Addressing these limitations necessitates exploring novel impeller designs that can overcome these challenges and improve bioreactor performance, by building on previous theoretical framework. Modified impeller designs, such as helical ribbon impellers and twisted blade turbines, have been studied for their enhanced mixing capabilities and reduced shear stress generation (Maryam *et al* 2020). A stirred bioreactor agitation system that consists of the popular Rushton turbine and new pitched blade impeller to study gas-liquid mass transfer was proposed previously by Gelves *et al* (2014). These impellers aim to achieve better flow patterns, improved mass transfer, and reduced energy consumption compared to traditional impeller designs. In another design, the impact of different impeller configurations, named Segment–Segment and Segment–Rushton, on impeller rotational speed and the hydrodynamic characteristics in a bioreactor, utilizing a dual impeller setup based on CFD was investigated (Ebrahimi *et al* 2019). Some comparisons were made between the two configurations, and the second seems to perform better. In addition, mixing characteristics in rectangular, octagonal and circular shapes of blades in a novel single configuration anchor impeller was studied in Kamla *et al* (2020), and it was realized that octagonal shape yielded the widest well-stirred region over the other cases. To further improve hydrodynamic responses in cell cultures in stirred bioreactor, a horizontal-dual bladed bioreactor for low shear stress was developed by Duman et al (2021). The findings suggest that the low shear horizontal bioreactor (LSB-R) design offers significant benefits in terms of reducing shear stress, mitigating excessive hydrodynamic forces, and promoting gentle operating conditions. Even in the recently modified and novel developed impellers for stirred bioreactors, there still exist concerns with optimum homogeneity and zero stagnant zone around the vessel walls.

In this work, we propose a flat blade vent-based impeller design for optimal homogeneity in bioreactors. The vent-based impeller incorporates auxiliary agitators in the form of vents parallel to the impeller blades, aiming to improve flow characteristics, enhance mixing efficiency, and promote better fluid circulation. The vents introduce axial and radial movements of the fluid particles, contributing to upstream dissipation and improving the overall flow patterns within the bioreactor. Hydrodynamic studies of stirred bioreactors based on impeller designs are typically carried out using experimental techniques such as particle image velocimetry (PIV), laser Doppler anemometry (LDA), and computational fluid dynamics (CFD) simulations (Ayodele *et al* 2022). These studies provide valuable data on flow patterns, velocity profiles, and shear stress distributions, enabling researchers to optimize impeller design, operating conditions, and scale-up processes. However, for optimization purpose, it is acceptable to use computational techniques that saves cost and time (Schirmer *et al* 2021, Oyejide et al 2021). In particular, CFD has gained relevance in biotechnology in the areas of scaffold modelling (Ebenezer *et al* 2021) and its fluid dynamics [Ali *et al* 2021], leading to better understanding of the bioreactor hydrodynamics and cell proliferation (Pong-Chol *et al* 2017, Bach *et al,* Vlaev *et al* 2020 2017, Alankar *et al* 2021). Therefore, the current study employed the CFD simulation approach, using $k$-$\epsilon$ standard model to investigate homogeneity, stagnant zones characteristics, and shear stress distribution in a lab-scale stirred tank bioreactors at speeds of 50, 100 and 150 RPM for both single and double impeller agitations. The CAD models, CFD models, mesh analysis, boundary conditions, results and presentations of computational variables such as velocity of upstream flow, impeller and shaft pressure, wall shear stress, eddy viscosity and turbulence eddy dissipation are presented in the subsequent sections.

**Nomenclature**

| | | |
|---|---|---|
| $T$ | - | Tank diameter |
| $ID_R$ | - | Internal domain radius |
| $ID_L$ | - | Internal domain length |
| $S_R$ | - | Shaft radius |
| $S_L$ | - | Shaft length |
| $C$ | - | Distance between base impeller and tank base |
| $U$ | - | Distance between top impeller and tank head |
| $H$ | - | Distance between the top and base impellers |
| $L$ | - | Total tank length |
| $D_S$ | - | Total distance between ends of blades |
| $B_W$ | - | Baffle width |
| $B_L$ | - | Baffle length |
| $IB_W$ | - | Impeller blade width |
| $IB_L$ | - | Impeller blade length |
| $\epsilon$ | - | Local energy dissipation rate |
| $V_{ol}$ | - | Vent hole length |
| $R_{vc}$ | - | Radius of vent chamfer |

**Abbreviations**

STBRs—Stirred tank bioreactors
*CFD*—Computational fluid dynamics
*WSS*—Wall shear stress
RANS—Reynolds-averaged Navier-Stocks
RPM—Revolution per minute

## 2.0 Methods
### 2.1 Geometry modelling
The CAD modelling in this study was performed on Solidworks software, and the conceptual drawing is presented in Fig. 1a and b, for two cases of a single and double impeller agitation configurations, respectively.

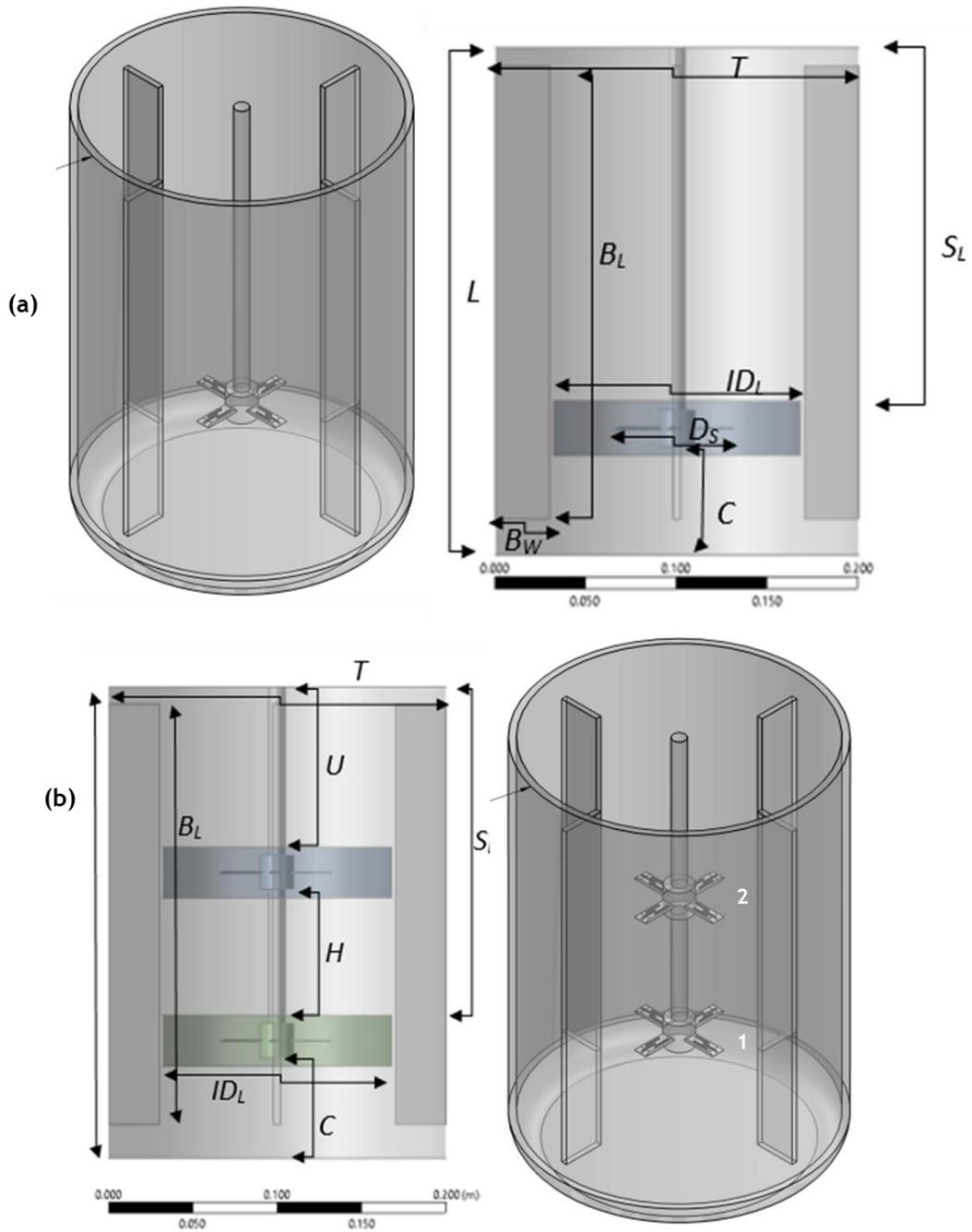

**Fig. 1.** 3D geometry of bioreactor setup in (a) single impeller and (b) double impeller agitation, with the top and base impeller named 2 and 1, respectively.

The vessels, shafts and impeller blades in both cases have same height, diameter and lengths of 280 mm, 210 mm and 25.29 mm. Other important parameters taken into consideration to develop the geometries are presented in Table 1 below.

Table 1. Parameters for developing the 3D models of the single and double impeller bioreactors

| Single impeller | | Double impeller | |
|---|---|---|---|
| Parameter | Value (mm) | Parameter | Value (mm) |
| T | 209.96 | T | 209.96 |
| $ID_R$ | 68 | $ID_{R1}$ | 68 |
| $ID_L$ | 30 | $ID_{L1}$ | 30 |
| $S_L$ | 210 | $S_L$ | 210 |
| - | - | $ID_{L2}$ | 30 |
| $S_R$ | 10° | $S_R$ | 10° |
| C | 40 | C | 40 |
| L | 280 | L | 280 |
| $D_S$ | 70.58 | $D_S$ | 70.58 |
| - | - | H | 80 |
| $B_W$ | 29.96 | $B_W$ | 29.96 |
| $B_L$ | 250 | $B_L$ | 250 |
| $IB_W$ | 10 | $IB_W$ | 10 |
| $IB_L$ | 25.29 | $IB_L$ | 25.29 |
| $V_{ol}$ | 11.35 | $V_{ol}$ | 11.35 |
| $V_{ol}$ | 11.35 | $B_L$ | 250 |
| $R_{Vc}$ | 0.30° | $R_{Vc}$ | 0.30° |
| - | - | U | 110 |

## 2.2 Vent design and bioreactor configuration

The vent geometries and bioreactor configuration were developed according to reference (Ebrahimi et al 2019). As presented in Fig. 1a and b, the proposed impeller comes with flat blades equally spaced into four cardinal positions around the impeller-shaft connector. On each of the blades are three parallel openings which are added to serve as auxiliary agitators during the main impeller rotation, strategically placed to facilitate the flow characteristics within the bioreactor. A zoomed image of the vents, and the designed flow directions is presented in Fig. 2b. The edges and vertexes of the blades were also designed to enhance traction during fluid flow across it, by fileting these regions. Studies in mechanical engineering have established that vent, such as air-vents, drain-waste vent system (Levin et al 2016) are of great significance in achieving optimal and undisruptive flow. In the case of a bioreactor, they can also provide an initial upstream dissipation around the impeller region right from the first rotation, thereby possibly leading to shorter time of nutrient delivery to the culture and even preventing the fluid particles from being trapped beneath the impeller blades.

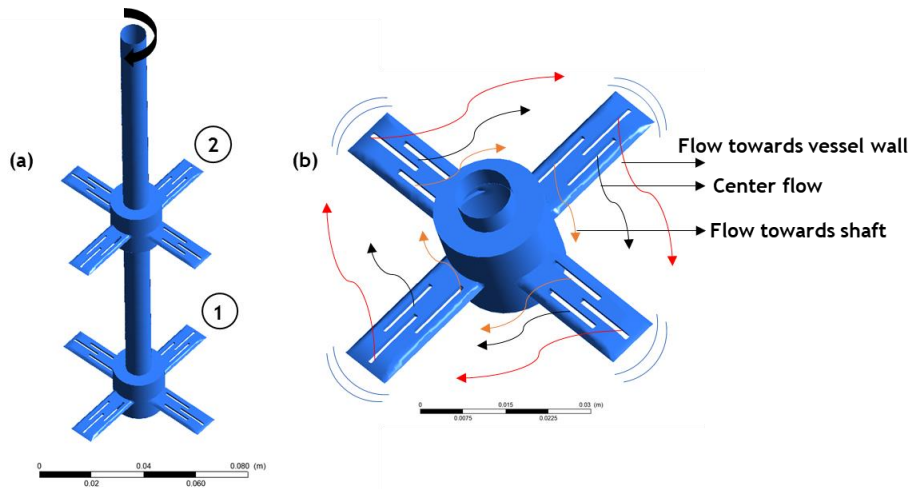

**Fig. 2.** (a) 3D model of double impeller configuration (80 mm apart) attached to the shaft and (b) the vent-based impeller detached from the shaft also showing the vent-imposed flow directions.

To configure the stirred bioreactor, only the impeller and baffles were imported from the Solidworks file into ANSYS Fluent design modular, where a primitive cylindrical shape was assigned to emblem the impeller connection and the necessary boundaries were assigned, such as the outer domain, inner domain(s), shaft and impeller. Stainless steel, glass and water were selected for the materials of the impeller, vessel wall and working fluid, respectively. According to studies, the driving power in laboratory scale bioreactors could range from 50RPM to 150RPM, depending on the height and stability of the shaft (Gelves *et al* 2014, Ebrahimi *et al* 2019). Hence, the three speeds were considered in this evaluation after preliminary the validation.

## 2.3   Mesh study

To solve the physics (presented in Subsec. 2.4), ensure accurate convergence, and present a physically realistic result, we used a mesh quality detection method that focuses on evaluating element skewness and smoothness. The skewness metric helps assess how much an element deviates from its ideal shape, with values closer to zero indicating a higher quality mesh (Sorgente *et al* 2023). The analysis was conducted using tetrahedral meshes of size 0.01 for both single and double configurations. The mesh consisted of 148,981 nodes and 806,403 elements for the single configuration, and 261,695 nodes with 1,415,880 elements for the double configuration. Quality metrics were categorized into bins representing various ranges, from excellent (0.0–0.5) to fair (0.6–1.0). Element counts within each quality range were recorded, revealing that the majority of elements fell within the 'excellent' categories. Fig. 3a shows the mesh of the models with the impeller and shaft buried in the vessel. Fig. 3b shows the mesh on the isolated impeller with the vents. The mesh fineness on the exposed shaft-impeller connection is shown in Fig. 3c and d for the single and double impellers, respectively.

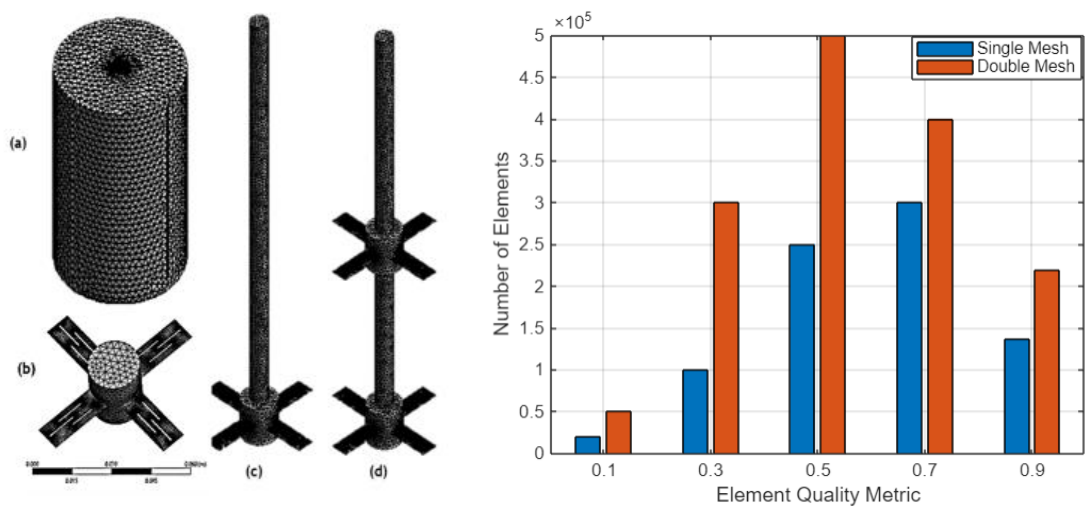

**Fig. 3.** Mesh fineness in entire bioreactor geometry (a), in isolated impeller with vents (b) and in shaft-impeller connection for single (c), and double configurations (d), respectively. (Right) Range of element metrics to number of elements to gauge the quality of the mesh in both single and double impeller agitation.

## 2.4.   Assumptions, CFD modelling and simulations

### 2.4.1.   Assumptions

The working fluid in the bioreactor is water, considered to be incompressible Newtonian fluid. Here, viscosity, density, thermal conductivity and specific gravity are assumed constant. The system is divided into two distinct regions: the stationary zone and the moving zone. The impeller is positioned within the moving zone, and its rotational speed is directly applied to this region. To ensure accurate modeling of the agitation process, the interface between the stationary and moving zones is characterized by a consistent surface mesh size. It is crucial that the interface meshes of both zones are identical in order to properly capture the dynamics of the agitation process. In addition, the fluid is assumed to represent a typical bioreactor fluid constituent for a small-scale bioreactor, so effects of oxygen, bubbles, and mass and heat transfer were not necessarily emphasized, as it is established that a general great mechanical agitation in a lab-scale bioreactor will enhance the above-mentioned constituents (Caillet *et al* 2021). Hence, this study mainly focuses on evaluating the effectiveness of the mechanical agitation of the proposed impeller.

### 2.4.2. CFD modeling and simulation

To model the hydrodynamic mixing in the stirred tank bioreactor, we discretized the governing partial differential equations using a finite volume method. The convection and diffusion terms were treated as follows:

The *convection term* was discretized using a first-order upwind scheme to ensure numerical stability. The discrete form for the utilized scalar variable $\emptyset$ is given by:

$$\left(\frac{\partial \phi}{\partial x}\right)_i \approx \frac{\emptyset_i - \emptyset_{i-1}}{\Delta x} \quad (1)$$

where and $\emptyset_i$ and $-\emptyset_{i-1}$ represent the values of $\phi$ at the current and previous grid points, respectively, and $\Delta x$ is the grid spacing.

The *diffusion term* was discretized using a second-order central difference scheme to achieve higher accuracy. The discrete form is given by:

$$\left(\frac{\partial^2 \phi}{\partial x^2}\right)_i \approx \frac{\emptyset_{i+1} - 2\emptyset_i + \emptyset_{i-1}}{\Delta x^2} \quad (2)$$

where $\emptyset_{i+1}, \emptyset_i$ and $\emptyset_{i-1}$ are the values of $\emptyset$ at the next, current, and previous grid points, respectively.

The numerical schemes were implemented using the ANSYS Fluent CFD software which employs these discretization techniques to solve the Navier-Stokes equations governing fluid flow and mixing.

Furthermore, the fluid is assumed Newtonian, with properties similar to water, hence Eq. 3 and 4:

$$\frac{\partial \rho}{\partial t} + \nabla \cdot (\rho u) = 0 \quad (3)$$

$$\frac{\partial (\rho u)}{\partial t} + \nabla \cdot (\rho u \otimes u) = -\nabla p + \nabla \cdot \tau + \rho g + \nabla \cdot \sigma \quad (4)$$

where $\rho$, $u$, $p$, $g$, $\tau$, and $\sigma$ are the fluid density, fluid average velocity, pressure, gravitational acceleration, viscose stress tensor, and Reynolds–Stress tensor, respectively.

In order to facilitate the process of mixing, we took into account the influences of impeller rotation, specifically the Coriolis and centrifugal forces, as well as the non-moving region. The governing equation for the region undergoing rotation and the region that remains stationary (Eq. 5 and 6), are expressed as follows:

$$\{\frac{\partial u_R}{\partial t} + \nabla \cdot (u_R \otimes u_I) = -\nabla p + \nabla \cdot \left(v_{eff}(\nabla u_I + (\nabla u_I)^T)\right) \quad (5)$$

$$\nabla \cdot u_R = 0$$

$$\{\frac{\partial u_I}{\partial t} + \nabla \cdot (u_I \otimes u_I) = -\nabla p + \nabla \cdot \left(v_{eff}(\nabla u_I + (\nabla u_I)^T)\right) \quad (6)$$

$$\nabla \cdot u_I = 0$$

where, the variables $u_I$ and $u_2$ represent the absolute velocities observed from the perspective of the stationary and rotating frames, respectively, measured in meters per second (m/s). The variable t represents the time in seconds (s), $v_{eff}$ denotes the effective kinematic viscosity in square meters per second (m²/s), and $p$ represents the pressure in Pascals (Pa). The boundary conditions are described by Eq. 7 and 8. In Eq. 8, the index $r$ represents the center of the elementary surfaces of each cell of the agitator. The variables $\omega, R, \Gamma 0, and\ \Gamma 1$ correspond to the angular velocity, impeller radius, vessel walls, and impeller walls, respectively. The rotational speed of the agitator is determined by the rotation speed of the mobile zone of the mesh.

$$No\ slip: u = 0, \qquad \frac{\partial}{\partial n}p = 0, \quad on\ \Gamma_0 \qquad (7)$$

$$Imposed\ velocity\ fields: v_r = \omega R, \qquad \frac{\partial}{\partial n}p = 0, \quad on\ \Gamma_1 \qquad (8)$$

For the shaft; specified as moving wall: rotational-axis (X, Y, Z) = (0,1, 0)

The Reynolds property for the mechanical agitation of the fluid is governed by: $R_e = \frac{\rho N d^2}{\mu}$

where $N$ is the impeller speed (rev/s), $\rho$ is the fluid density (kg/m³), $d$ is the impeller diameter (m) and $\mu$ is the dynamic viscosity (Pas).

### 2.4.3 CFD simulations

Once the mesh was generated, boundary conditions were fixed and CFD parameters assigned, the numerical simulation was performed. To setup the solution, Pressure-Velocity Coupling scheme was employed using SIMPLEC. Standard initialization method was used to compute the simulation from all zones for 5-time steps at time step size of 0.01, at a reporting interval of 1. Pressure-correction under relaxation factor is generally set to 1.0 (Liu *et al* 2022), aiding in the convergence speed. The continuity solution reached a convergence at e-02, e-03 and e-03 in both impeller configurations at speeds of 50, 100 and 150 RPM, respectively. The post-processed results where the velocities of flow, associated pressure, velocity vectors at the vents, shaft and impeller pressure and wall shear stress, and eddy dissipation behaviors.

### 3.0    Results and discussion

Herein, six (6) separate simulations, and a total of 24 in all, were performed for the flow velocity, pressure, velocity vector and wall shear stress in both the single and double impeller configurations at speeds of 50, 100 and 150 RPM, described in case a, b and c, respectively. Information on the contours and legends were further depicted in graphs, taking values of such characteristics (for instance, velocity) from beneath the rotating impellers and particularly around the region of the four impeller blades. In this section, t and b are used in the legends of the graphs to indicate top impeller and bottom impeller in the double configuration.

## 3.1 Results
### 3.1.1. Velocity contours, vectors and streamline in single impeller agitation

The contours, vectors and streamline of the fluid velocity on the XY Plane at Z=1, are represented in Fig. 4 a-f at speeds of 50 RPM, 100 RPM and 150 RPM, respectively. The main factors of interest were the flow movements and concentration within the tank vessel.

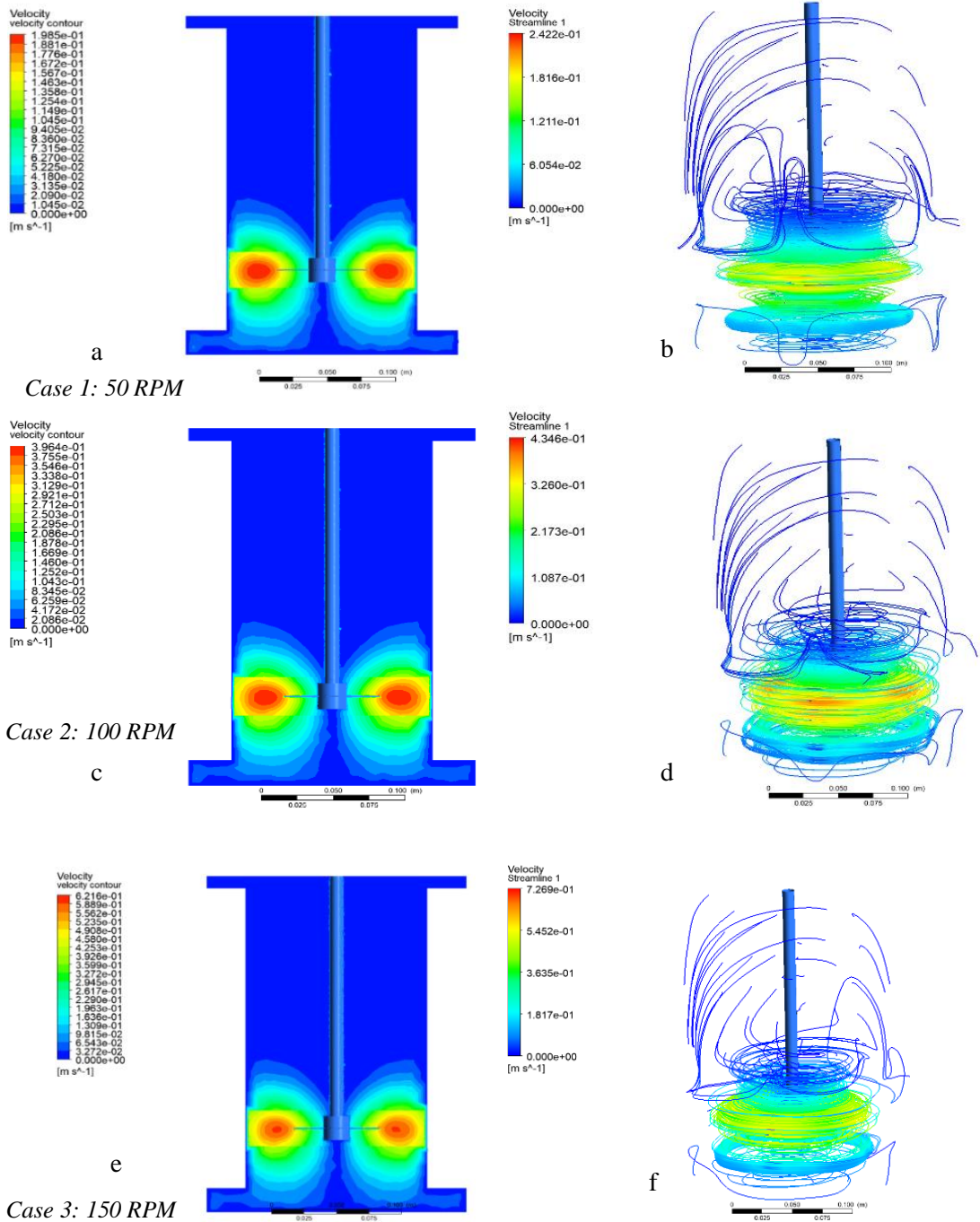

a  
b  
*Case 1: 50 RPM*

c  
d  
*Case 2: 100 RPM*

e  
f  
*Case 3: 150 RPM*

**Fig. 4.** Velocity contour and streamline for single layer impeller at 50, 100 and 150 RPM in Case 1(*a*, *b*), Case 2 (*c*, *d*) and Case 3 (*e*, *f*), respectively.

The result in Fig. 4 a-b, c-d, and e-f show a mirrored flow behavior on the two halves of the shaft, with the highest flow magnitude observed at the extreme end of the impeller where its blades cut through the fluid. As the impeller rotates, the fluid could be seen agitated upstream with a velocity that drastically dropped from about 1.881 e-01 to 2.090e-02, 3.755e-01 to 2.088e-02 and 5.88e-01 to 3.272e-02 for speeds of 50, 100 and 150 RPM, respectively. In the three cases, stagnant zones were significantly mitigated, as the fluid could be seen to propagate towards the walls of the vessels in axial direction, and slightly touching it. A graphical representation of the flow velocities in the three cases is presented alongside the double impeller configuration in Fig. 7, in which readings were taken right below the impeller at Points 1: x=-0.15, y=-0.23, z=0 and Point 2: x=0.15, y=-0.23, z=0 and a straight line. It can be seen from the graph how uniform the velocity of flow is, right from the center of rotation (in x-direction) and away towards the entire circumference of the rotating impeller blades.

The streamlines, presented in Fig. 4 b, d and f, show 3D details of the hydrodynamic behavior, particularly a realistic pattern in which the agitation would move the fluid particles in the vessel towards the placed cell culture. Due to the turbulent nature of the flow, vortices rings were seen at the base of the impellers for the three speeds under consideration. The thickness is more pronounced at speed of 50, 100 and 150 RMP, respectively. With respect to dissipation, there was distinct circulatory flow pattern in the three cases. In case 1, with speed of 50 RPM, the greatest dissipation was seen about mid-way up the impeller, with tiny lines of fluid shooting up the vessel's height. However, the rotation was concentrated towards the center of the vessel, expect at the region of the impeller blades, where the fluid closes up with the vessel walls. With an increasing speed of 100 RMP, the rotated fluid occupies more space in the vessel, both at the base and around the rotating impeller. As in case 1, tiny lines of fluid also moved upward to the roof of the vessel in the same pattern. Although there are similarities between the dissipation in case 2 and 3, at 150 RMP, the fluid loses concentration towards the center along the shaft, and an obvious space was observed between the vortex ring and the oscillating fluid concentration.

The velocity vector is presented in Fig. 5 (case 1, 2 and 3). Here, the auxiliary agitators (the vents), are seen to contribute to upstream dissipation around the impeller region as the blades cut the fluid in opposite direction to the axis of rotation of the shaft. This shows that the vent, though designed horizontal to the shaft, contributes to rotational movements of the fluid particles. In addition, it was observed that this behavior was similar for the three cases at speeds of 50, 100 and 150RPM, respectively; and the velocity of flow across the three vent openings was greater at speed of 150, 100 and 50 rpm, respectively.

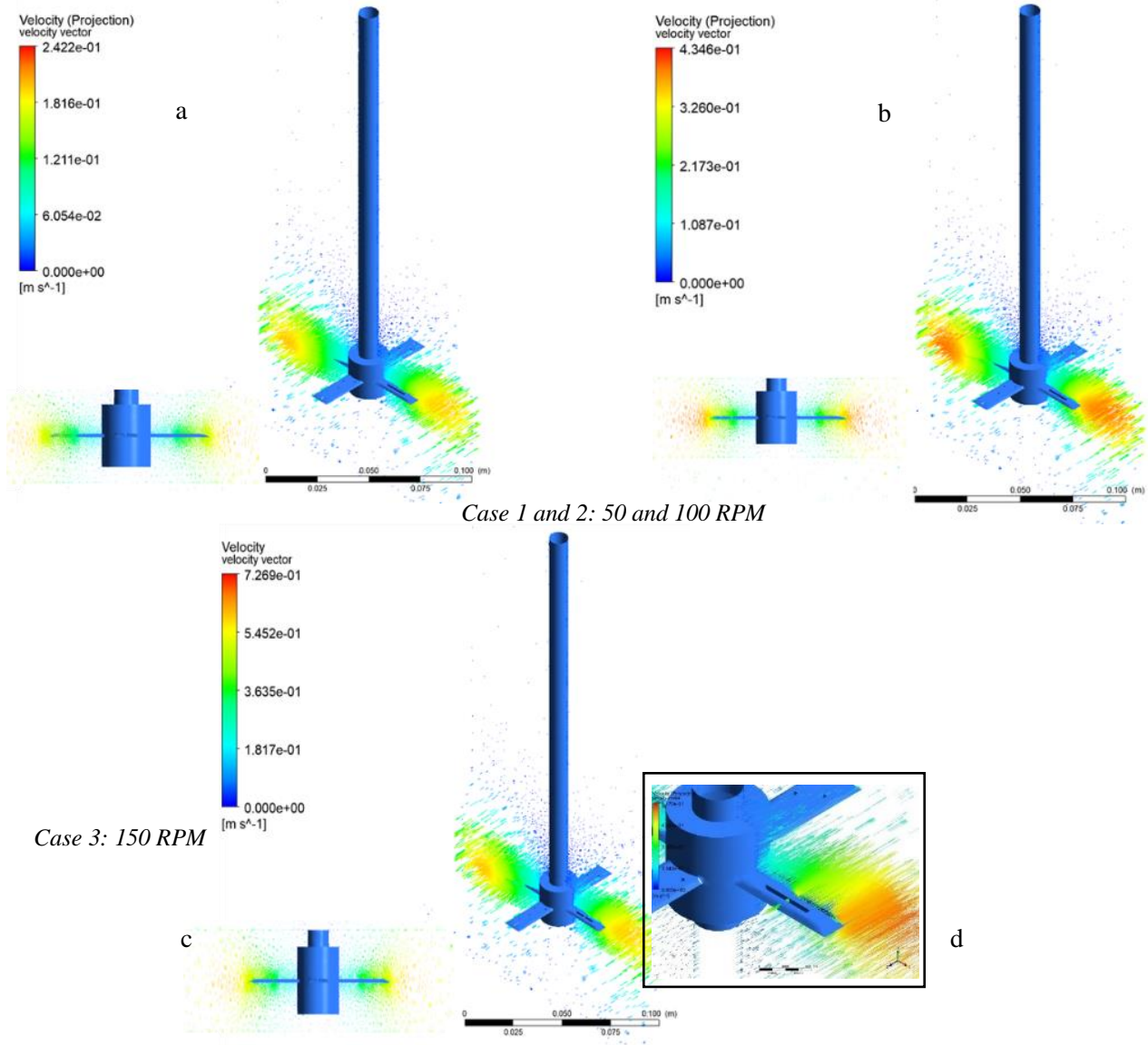

**Fig. 5.** Velocity vector for single impeller at speed of 50 and 100 RPM in Case 1 and 2 (a & b) and 150 RPM in Case 3(c), respectively. (d) Expanded view of the velocity vector particularly through the vents, at 150 RPM.

### 3.1.2. Velocity contours, vectors and streamline in double impeller agitation

Similar to the results for the single impeller configuration, the velocity contours in the dual configuration were generated on the XY Plane at Z=1 at speeds of 50 RPM, 100 RPM and 150 RPM as shown in Fig. 6.

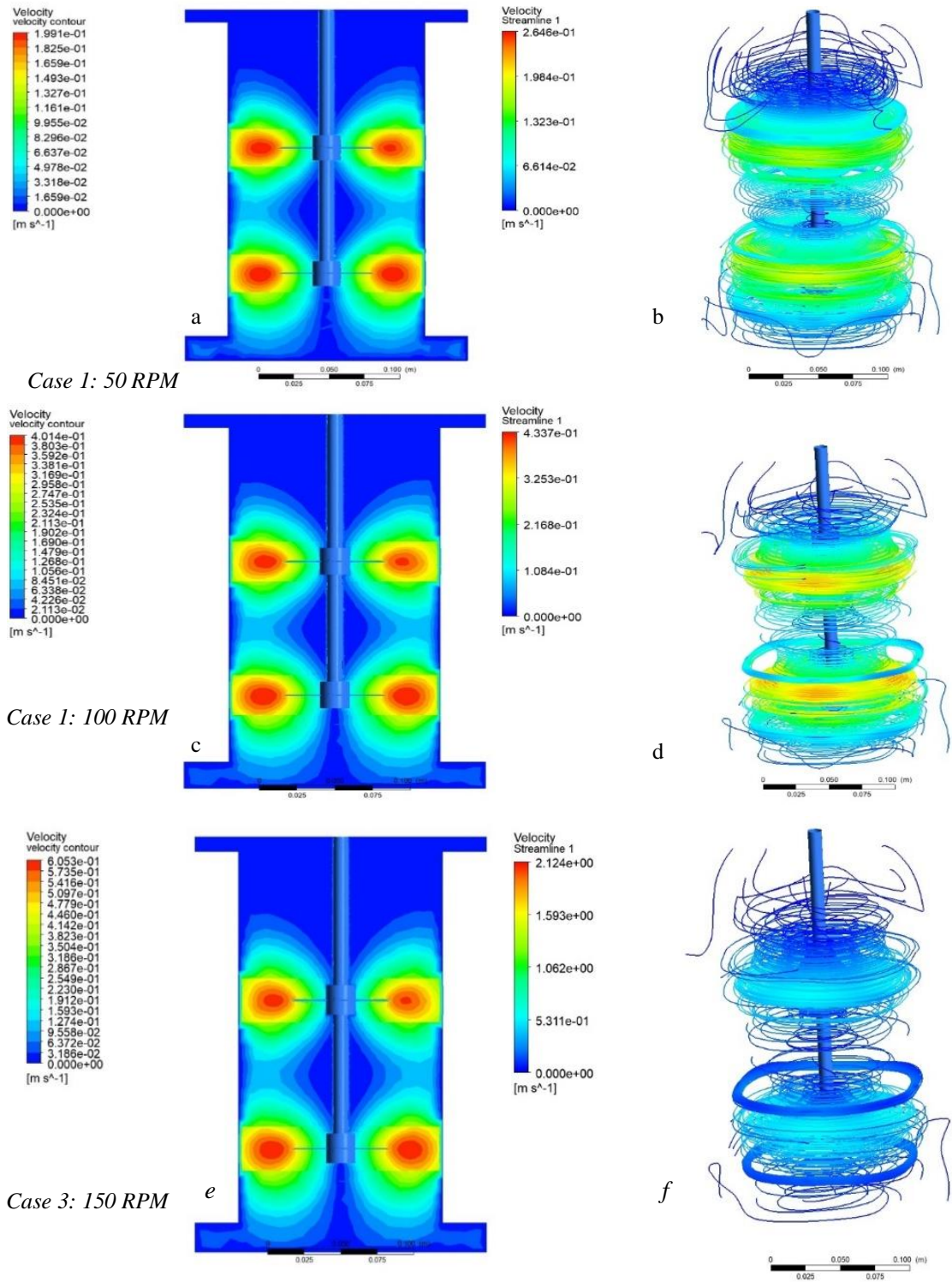

*Case 1: 50 RPM*

*Case 1: 100 RPM*

*Case 3: 150 RPM*

**Fig. 6.** Velocity contour and streamline for double impeller configuration at speeds of 50, 100 and 150 RPM in case 1(a, b), Case 2 (c, d) and Case 3 (e, f), respectively.

In Fig. 6, case 1, it can be seen that at a speed of 50 RPM the velocity of flow has the highest magnitude at the immediate surroundings of the tips of the top and base impeller blades. The flow quickly develops towards the shaft and center of rotation of the impellers, and the particles of the developed flow merge, moving upstream with an intermediate intersection without leaving a space between due to the similar flow patterns interjecting each other. With an increase in speed from 50 RPM to 100 and 150 RPM (Fig. 6, case 2 & 3), the velocity at the ends of the impeller increases from 1.991 exp-01 to 4.014 exp-01 to 6.053 exp-01, respectively. It is also noted that the velocity at the top impellers is more concentrated than the base impellers, which can be caused by the depth, as observed in most submerged objects. The three cases also show significant reduction in stagnant zone, which is obviously seen to be less than 1% of the entire area of the bioreactor vessel.

For the streamlines, presented in Fig. 6b, d & f, at a speed of 50, 100 and 150 RPM, respectively, the flow distribution pattern is similar, though there were obvious and significant implications of the slight differences. At 50 rpm, the upstream fluid particle dissipation across the vessel through the base and top agitation, are at close proximities to each other, closing up the entire region of culture placement, with rich homogeneity. The streamline also shows that the stagnant zone in the vessel at a speed of 50 RPM is negligible, as it occurs slightly towards the walls of the space between the impeller but not around the culture. The same could be seen for the increased speed of 100 RPM (Fig. 6d) but with less homogeneity and slightly wider stagnant zone around the culture and the walls of the bioreactor vessel. There is also the presence of vortex ring layers, beneath and above the base impeller, which was not pronounced at the speed of 50 RPM. However, at 150 RPM, a form of tightly packed vortex was seen at the base which could be characterized by particle condensation, probably due to the high-speed force exerted downwards during the rotation of the impellers. Importantly, it is realized that the flow pattern in the three cases was axial, which falls in line with most flow patterns found in bioreactor impellers.

The flow velocity behavior is further described in Fig. 7 using line graphs. The readings were taken right below the top and base impellers at Points 1: x=-0.15, y=-0.13, z=0 & Point 2: x=0.15, y=-0.13, z=0, and also Points 1: x=-0.15, y=-0.23, z=0 & Point 2: x=0.15, y=-0.23, z=0 for speeds of 50, 100 and 150 RPM. It is observed that as the velocities increase, the plot pattern was maintained, though the velocity for the base impeller has a higher velocity range compared to the top impeller.

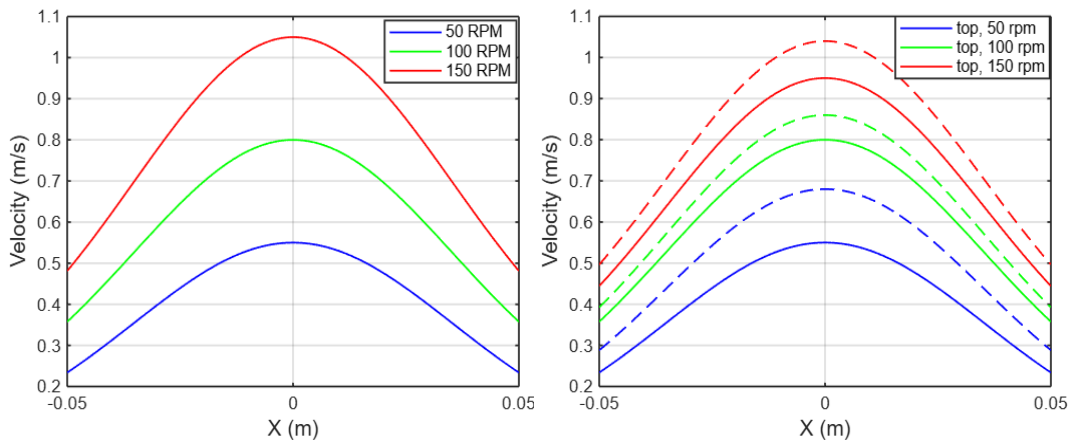

**Fig. 7.** Velocity graph. (left) single impeller (right) double impeller configuration at the various speeds on the top and base impellers

Furthermore, the velocity vector for the double impeller is presented in Fig. 8 a-c, at speed of 50, 100 and 150 RPM. Similar to the single impeller, a close look at the impeller velocity vector shows flow through the holes of the vents while also following the general direction of the overall movement. It was also noted that the base impeller generated the highest velocity, implying that it was doing most of the upstream work to the top impeller.

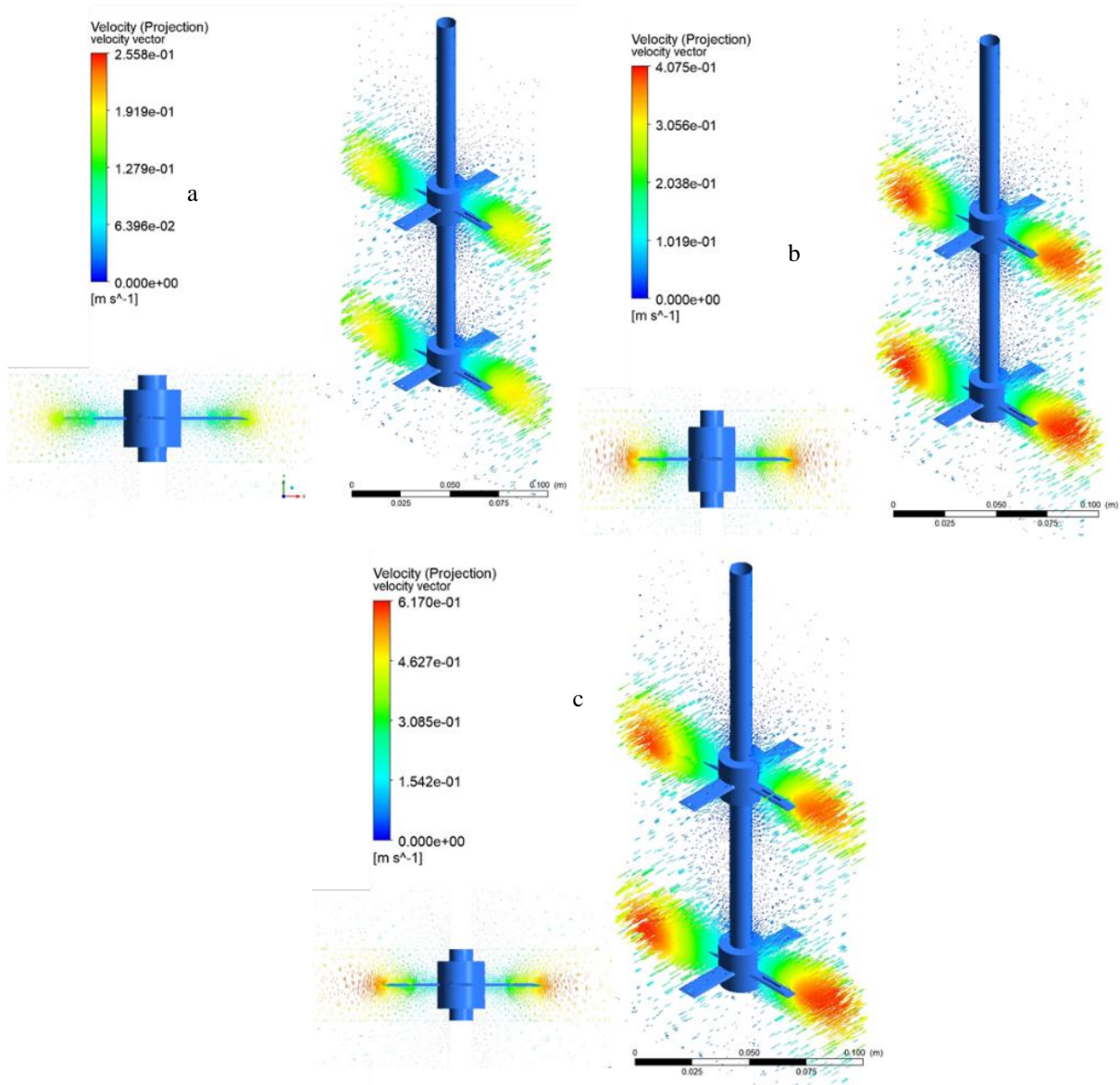

**Fig. 8**. Velocity vector for double layer impeller at speed of 50, 100 and 150 RPM in Case a, b and c, respectively.

### 3.1.3 Pressure contour in single impeller agitation

The vessel pressure contours for the single impeller configuration, generated on the XY Plane at Z=0, and at speeds of 50 RPM, 100 RPM and 150 RPM are presented in Fig. 9 a-c. In the three cases, the pressure distribution behavior was distinct. First, at speed of 50 RPM, relatively high (about 3.147exp+00) and uniform pressure was observed within the vessel, expect downstream and mid-up stream around the circumference of the rotating impeller where the pressure was lesser (-1.778exp+01). Second, at an increased speed of 100 RPM, pressure within the vessel become higher (about 3.602exp+00), except at the impeller, where it drops to -4.492exp+01. The pressure distribution in the second case was similar for case 3 at 150 RPM. However, the vessel pressure was highest in case 3 (about 5.382exp+00), which dropped to about -9.213exp+01. When compared, pressure dropped significantly around the impellers in the three cases, which could be confirmed from the streamlines in Fig. 4 b and d. In addition, the pressure drop was more pronounced as the speed increased from 50 to 100 and 150RPM in term of visual observation, but the highest magnitude was recorded at speeds of 150, 100 and 50, respectively.

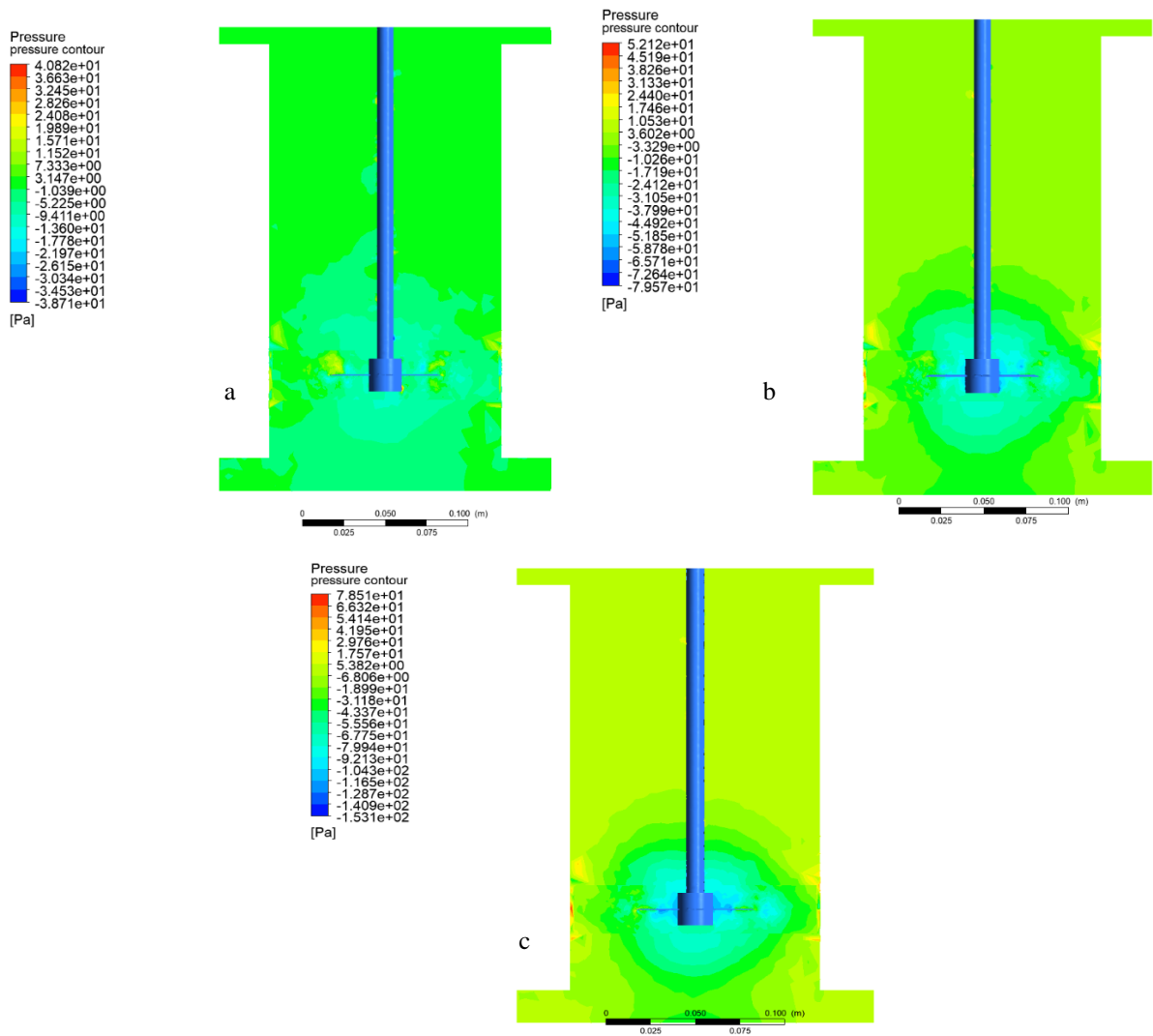

**Fig. 9.** Pressure contour of the single impeller configuration for Case 1(a), Case 2(b) and Case 3(c) at speeds of 50, 100 and 150 RPM.

A graphical representation of the pressure distribution around the impellers along the X-axis, taken at the same point as in the velocity graphs for the various speeds is shown alongside the double configuration in Fig. 11. It can be observed that the higher the speed applied, the lower the pressure recorded on the graph.

### 3.1.4 Pressure contour in double impeller agitation

As expected, the result for pressure distribution in the dual impeller bioreactor was quite different from that of the single impeller bioreactor at the three speeds, shown in Fig. 12 a-c. As a consequence of the agitation, irregular distribution was observed at 50 RPM (Fig 12a), with some sort of high pressure at the teeth of the base and top impellers. Unlike in the single impeller bioreactor at speed of 50 RPM, pressure did not significantly drop around the impeller; just a little around the base impeller (about -2.300exp+01) but almost nothing around the top impeller. In case b, at a speed of 100RPM, the contour reveals significant pressure drop (about -4.278exp+01), that engulfs both impellers and propagates towards the vessel walls. In case *c*, at 150 RPM, high pressure distribution was pronounced in the entire vessel (4.528e+00), except at the impellers where it drops to about -7.739e+01. This effect could have resulted from the intensive speed of the impeller.

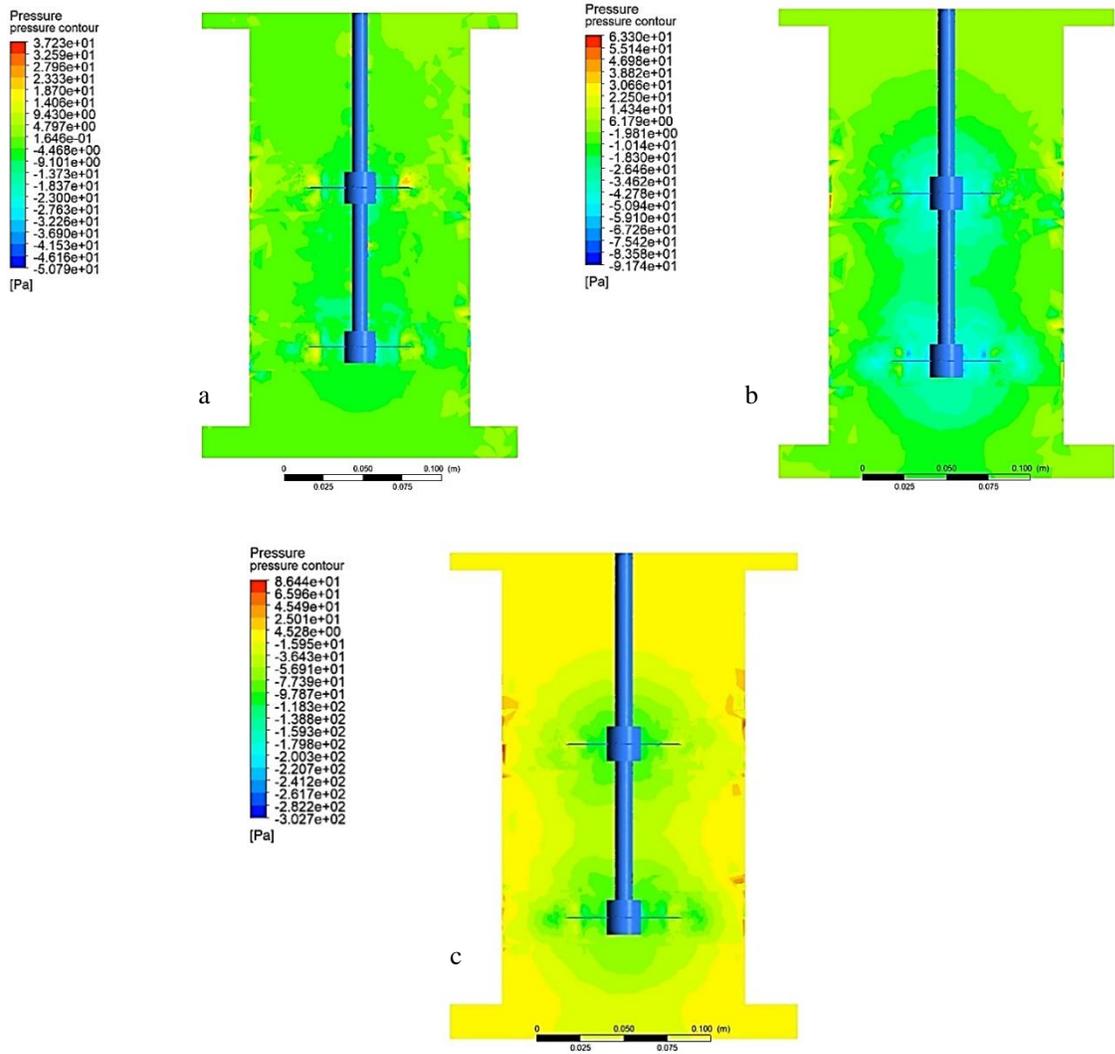

**Fig. 10.** Pressure Contour of the double impeller configuration at speeds of 50, 100 and 150 RPM for Case (a), Case (b) and Case (c), respectively.

The pressure distribution around the impellers alone, is further presented in Fig. 11, using the same coordinates as the double impeller velocity plots. It can be seen that for the top impeller, the ending points of the pressure graph are at a closer range compared to the base impeller in which it is much more spaced, and the pressure magnitudes are lower than in the top impeller.

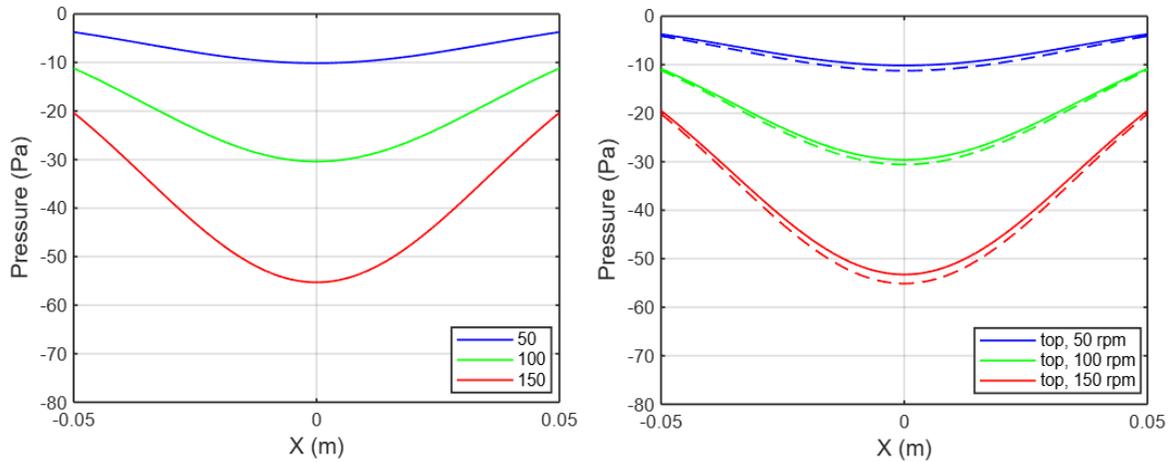

Fig. 11. Pressure graph. (left) for single impeller and (right) double impeller configuration at the various speeds on the top and base impellers.

### 3.1.5 Impeller wall and shaft pressure in single impeller agitation

Pressure on the impeller and shaft walls are presented in Fig. 12 a-c for 50, 100 and 150 RPM, respectively, and the effect is emphasized by isolating the impeller in the side views. The result for the three speeds were different, though the pressure pattern on the blades follows the same theme in each case. In case *a,* higher pressure was seen at the fluid cutting vertices of the four blades. Pressure on the shaft was lesser than on the impeller blades, but the lowest pressure was observed around the rim (about 1.091e+02) connecting the impeller to the shaft. As the speed increases to 150 RPM, lesser pressure was observed on the impeller blades and it was evenly shared by the impeller blades. The pressure dropped significantly (-3.976+01), too, at the connection, but the shaft pressure was higher than found at speed of 50 RPM, at about 1.922e+00 to 2190e+00. The observations were exaggerated in case *c*, at increased speed of 150RMP. The shaft pressure was 8.697e+00, and pressure at the impeller-shaft connection dropped to about -9.238+01. Obviously, pressure on the impellers reduces as speed increases.

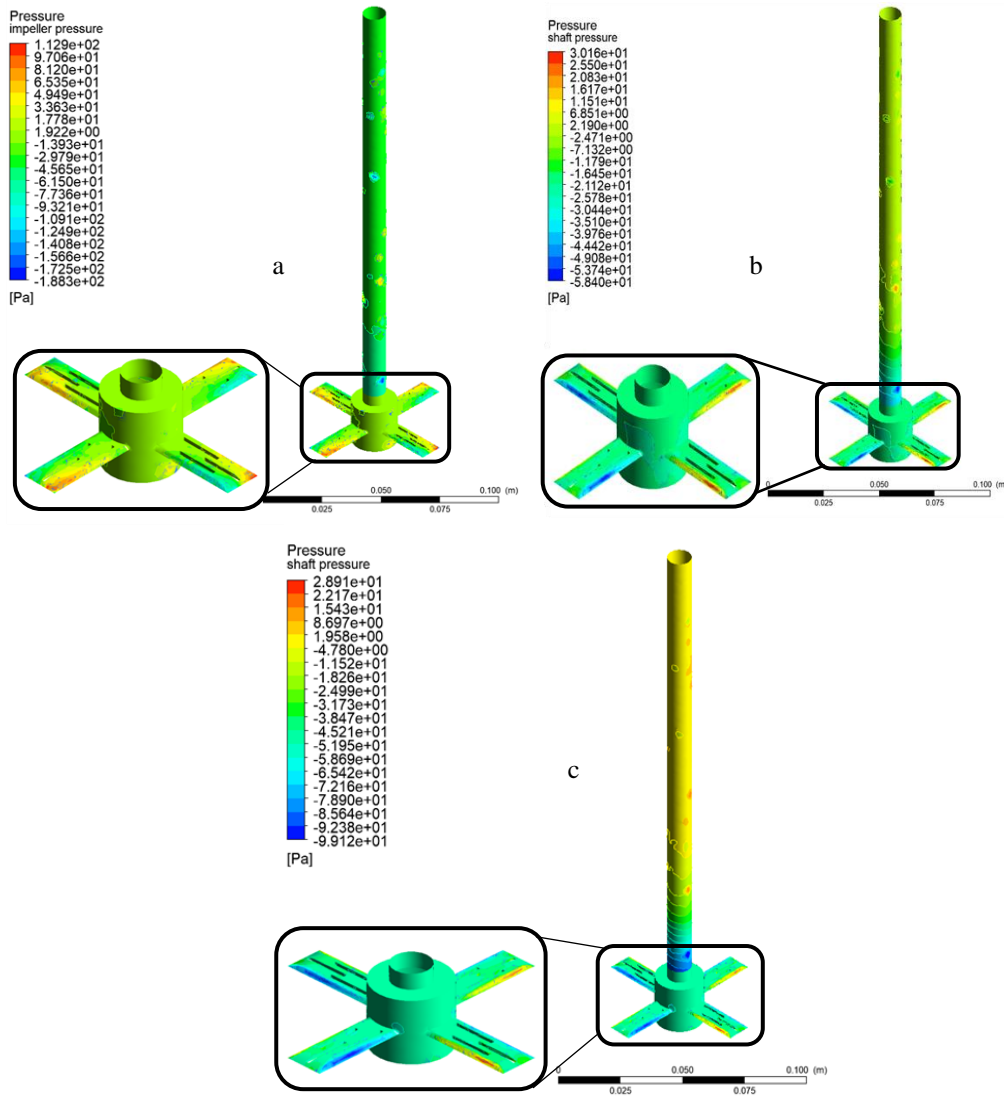

**Fig. 12.** Wall pressure on the single impeller and shaft at speeds of 50, 100 and 150 RPM, in Case (a), Case (b), and Case (c), respectively

### 3.1.6 Impeller wall and shaft pressure in double impeller agitation

The impeller and shaft pressure for the dual impeller agitator is presented in Fig. 13 a-c at speeds of 50, 100 and 150 RPM. Again, as in the case of the single impeller agitation, the pressure on the impellers in these cases reduce as speed increase. However, a notable observation was the increasing high pressure distribution mid-way upstream, which was about 3.574e+00, 7.392e-01 and 1.790e+02 at 50, 100 and 150 RPM, respectively. Also, there was almost no pressure drop at the connecting circumference of the impeller and shaft in case *a* and *b*, but a significant drop was observed beneath and above the region of connection in case *c,* in both the top and base impellers.

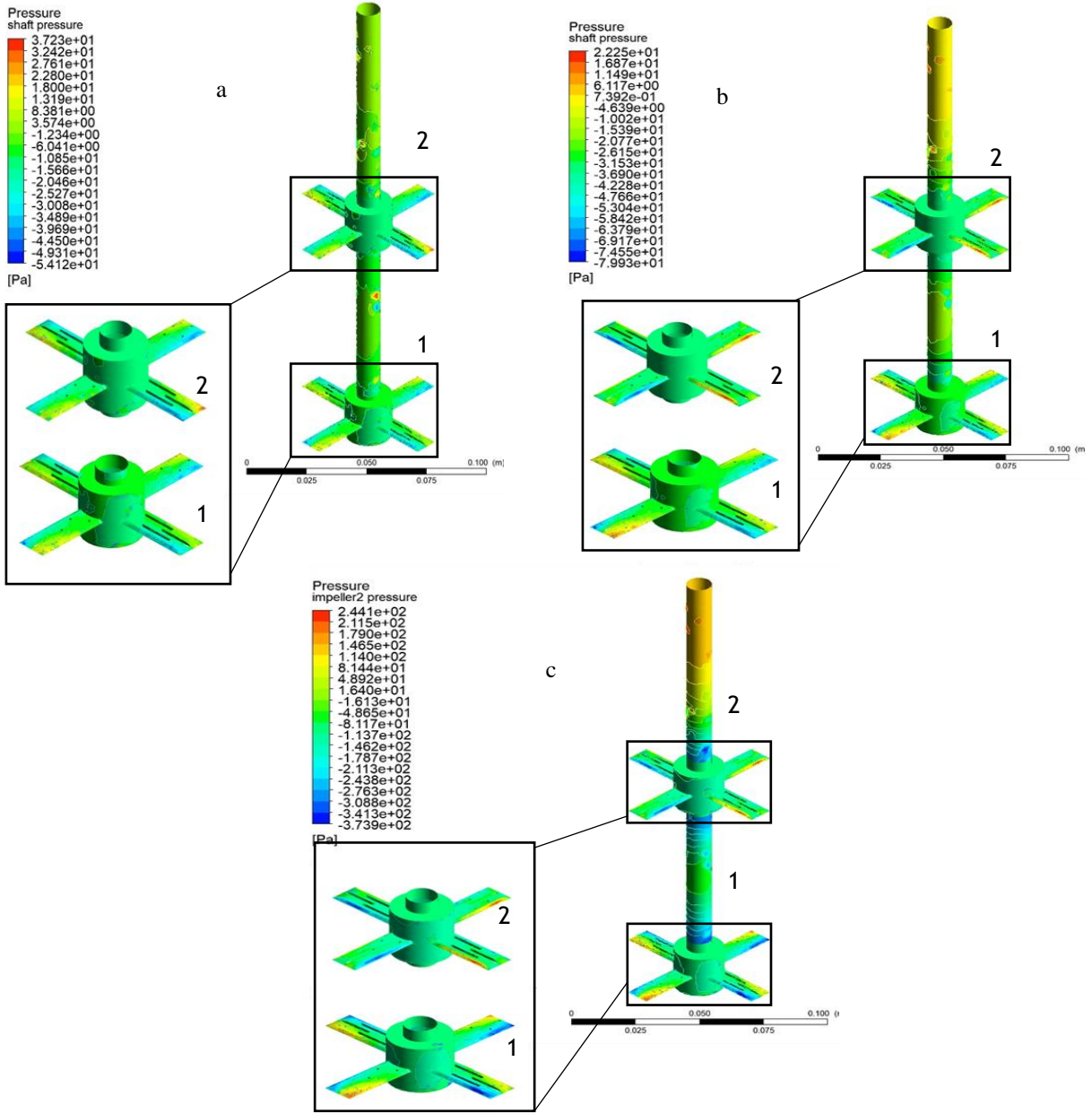

**Fig. 13.** Wall pressure of the double layer impeller and shaft at 50, 100 and 150 RPM for Case (a), Case (b), and Case (c), respectively.

### 3.1.7   Wall shear stress in single impeller agitation

The wall shear stress on the shaft and impellers are also presented in Fig. 14 a-c for the single impeller agitation setting. Here, it is noticed that the wall shear increases around the areas of the impellers as the speed increase. The values of the maximum WSS on the impeller blades are about; 5.128e-01Pa, 6.588e+00Pa, and 8.388e-01Pa, while the maximum WSS on the shafts are about 5.985e-01 Pa, 1.975e+01 Pa and 2.309e+00 Pa at speeds of 50, 100 and 150 RPM, respectively. Notably, the wall shear stress is mostly concentrated around the edges/tips of the impeller blades, which is the major continuous contact point with the fluid as the shaft rotates. The WSS in the three cases is however not as prominent, as the vents create a form of flow distribution mechanism and light weight effect. The design of the blade width also helps gliding, thereby reducing the severity of the shear stress on the impeller. Furthermore, the entire wall of the shafts (creating the main rotary force) in the three cases were affected, which is due to high frictional tangential force between its surface and the turbulent fluid flow.

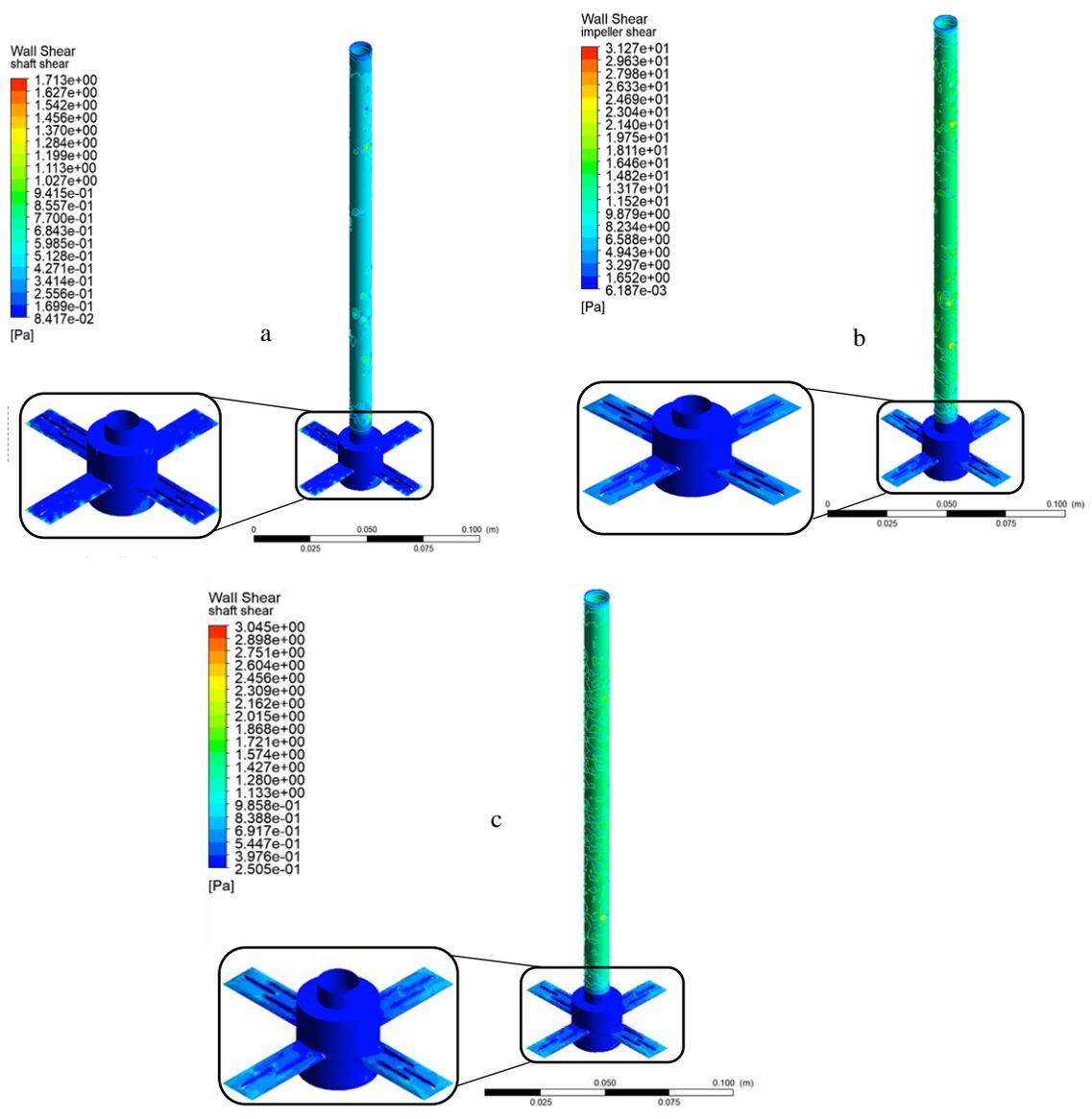

**Fig. 14.** Wall shear of the single layer impeller and shaft at speed of 50, 100 and 150 RPM for Case (a), Case (b), and Case (c), respectively.

### 3.1.8   Wall shear stress in double impeller agitation

There is a close relationship between the WSS results in both the single and dual impeller agitation settings. Here too, the WSS on both the impeller and shafts increases as speed increases (Fig. 15 a-c). However, when observing the simulation outcomes on the top and base impellers, it is obvious that the wall shear is heaviest on the top impellers, which seems to be negligible at speed of 50RPM, but heavy at speeds of 100 and 150 RPM with values of 3.845e+00Pa and 4.143e+00Pa, respectively. This difference in WSS behavior in the top and base impellers is perceived to be due to the workload on the top impeller as the base impeller displaces the fluid upstream against its walls, sequel to the actual WSS generated from the flow across its blades during rotation.

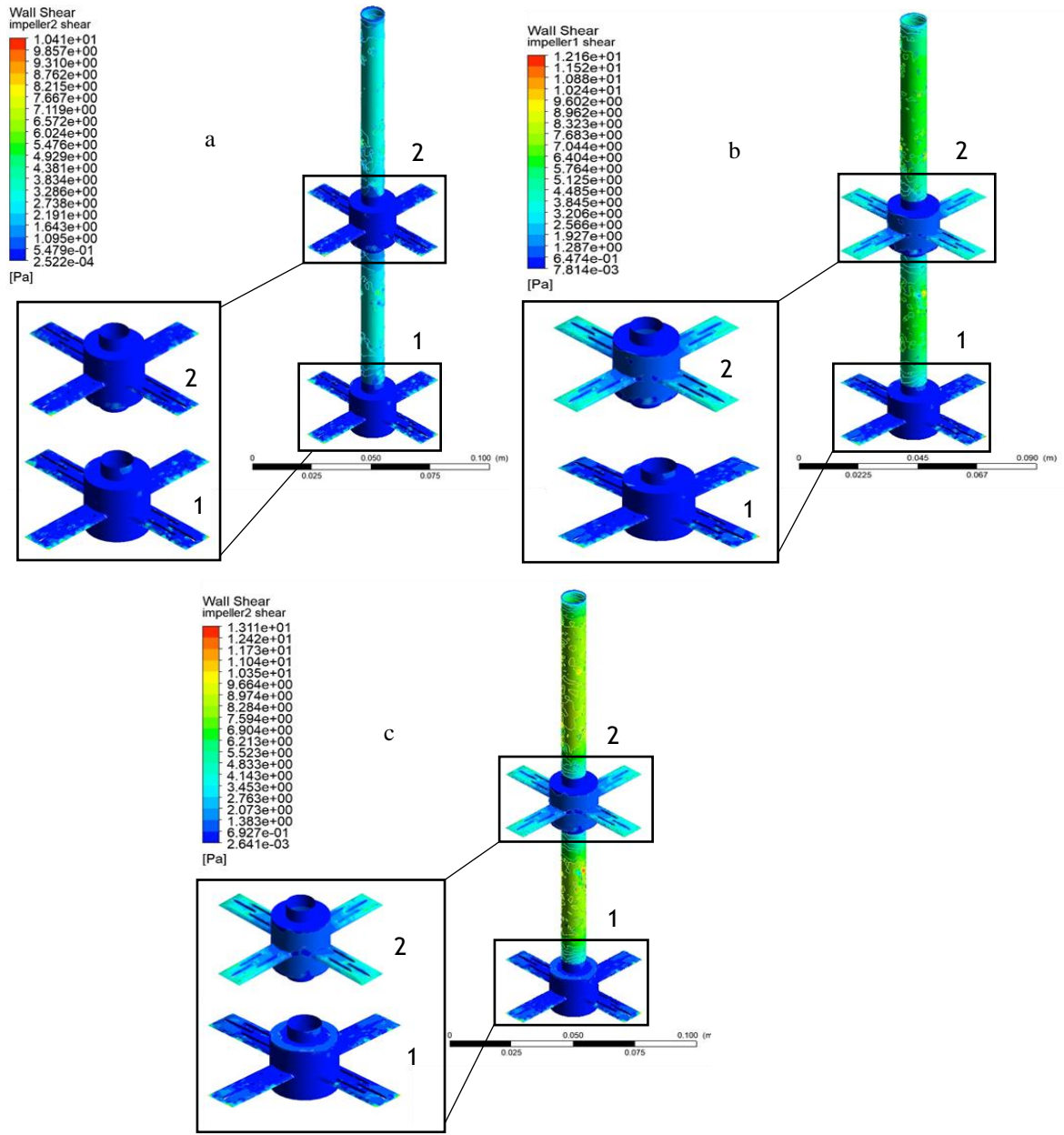

**Fig. 15.** Wall shear of the single layer impeller and shaft at speed of 50, 100 and 150 RPM for Case (a), Case (b) and Case (c), respectively.

### 3.1.9 Eddy viscosity, turbulence eddy dissipation and turbulence kinetic energy

In the streamlines in Fig. 4 and 6 d, e, and f, different turbulence behaviors were observed at the different speeds, with all cases involving some kind of vortex rings. Although heat and mass transfers were not modelled in the current work with some justifications earlier mentioned in the assumptions, it is essential to predict the major turbulence parameters in the flow. Hence, the eddy viscosity, turbulence eddy dissipation and turbulence kinetic energy were studied as a property of flow and turbulence model. Fig. 16 shows the plots for eddy viscosity, turbulence eddy dissipation and turbulence kinetic energy for the single and double impeller configurations based on the turbulence model.

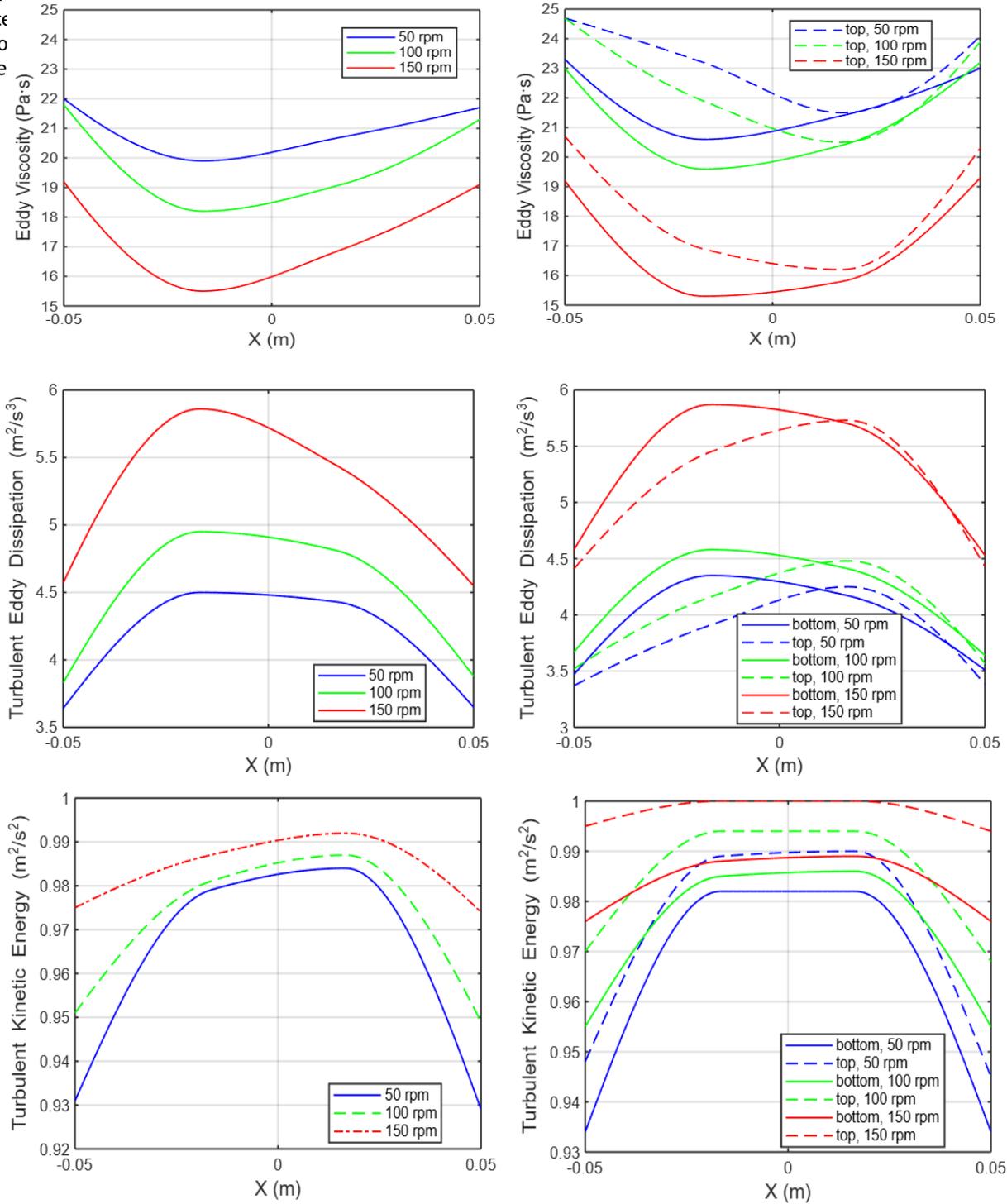

**Fig. 16.** Comparison between eddy viscosity, turbulence eddy dissipation and turbulence kinetic energy in single (left) and double impeller (right) configurations at speeds of 50, 100 and 150 RPM, respectively.

It can be seen that for the single impeller configuration (Fig. 16a; left), the eddy viscosity reduced with an increase in speed and at speeds of 50 and 100 RPM, the values almost reach the same ranges. However, in the double impeller configuration, the eddy viscosity differs slightly for the top and base impellers agitations, where the base impeller has lower values of eddy viscosity and the speed variants have a closer relational value to each other (Fig. 16a; right). The graph suggests that the highest eddy viscosity was reached in both configurations. It is established that with the presence of eddies, there is a chance for turbulence energy dissipation rate Turizo *et al* (2021), a phenomenon that can also help in identifying the level of intensity of turbulence gotten in the agitation processes. Here (Fig. 16b), it is observed that the turbulence eddy dissipation in both the single and double impeller configurations have similar behavior with the eddy viscosity pattern in both cases at the three speeds (50, 100 and 150 RPM), except that it occurred in an upward-down manner instead of downward-up, as in the eddy viscosity. Both impeller configurations also displayed a uniform pattern of the eddy dissipation, though relatively low in the single configuration (Fig. 16; left). And for the double impeller (Fig. 16b; right), the eddy dissipation creeps up more uniquely at the upper impeller compared to the base impeller and is higher as speed increases. Furthermore, the accompanying kinetic energy, as seen in Fig. 16c, increases with speed, though the magnitude at the center of rotation at 50 and 100 RPM relatively coincide while a distinct pattern was observed at speed of 150 RPM. In the double-impeller configuration (Fig 16c; right), the turbulence kinetic energy pattern was entirely different for both the top and base impellers, forming similar increasing convex patterns at speeds of 50, 100 and 150 RPM, respectively in the base agitation. However, for the top impeller, the turbulence kinetic energy pattern was somewhat uniformly straight at speeds of 100 and 150 RPM, but produce a convex shape; as seen in the base impeller; at speed of 50 RPM.

## 4.0   DISCUSSIONS

### 4.1. Velocity, streamlines and vector of flow in single and double impeller agitators

The results presented in Fig. 4, 6, 7, 11 and 16 for both the single impeller and double impeller agitation reveal important implications about the performance of the novel impeller design on the hydrodynamic behavior in the stirred-bioreactor. Firstly, the observed mirrored flow behavior on the two halves of the shaft in both cases, with the highest flow magnitude at the extreme end of the impeller, is consistent with the concept of fluid dynamics in stirred tanks. It is well known that the impeller blades cutting through the fluid induce a strong flow in their vicinity. This phenomenon leads to increased fluid agitation and mixing near the impeller, which is crucial for promoting efficient mass transfer and homogeneity within the bioreactor (Lins *et al* 2022). Moreover, the drastic drop in velocity as the impeller rotates further supports the effectiveness of the vent-based impellers in generating fluid motion. The reduction in velocity observed at different rotational speeds (50, 100 and 150 RPM) aligns with the established relationship between impeller speed and flow intensity. Higher rotational speeds lead to more vigorous agitation, resulting in increased fluid velocity near the impeller blades (Singh *et al* 2021).

The uniformity of flow velocity throughout the bioreactor, as demonstrated by the graphical representation in Fig. 7, 11 and 16, is a desirable characteristic in bioprocessing. The ability to maintain a consistent flow velocity from the center of rotation to the circumference of the impeller blades ensures more uniform mixing and distribution of nutrients, gases, and cells throughout the culture medium (Ebrahimi *et al* 2019). This uniformity facilitates optimal conditions for cell growth, product configuration, and nutrient uptake, ultimately improving bioprocess efficiency. Furthermore, the streamlines shown in Fig. 4 and 6 b, d, and f, provide a 3D visualization of the hydrodynamic behavior in the bioreactor. The observed movement of fluid particles towards the placed cell culture demonstrates the potential for effective transport and mixing of the desired components. In addition, the configuration of vortex rings at the base of the impellers, especially at higher speeds and in the double impeller configuration, further enhances fluid circulation and mixing, thereby reducing stagnant zones (see Fig. 17). Vortices are known to contribute to improved mass transfer rates and enhanced mixing by promoting better fluid dispersion and reducing the presence of stagnant zones (Li *et al* 2022).

One notable difference between the hydrodynamics of the single and double impeller agitation, as observed from the velocity contours and streamlines, is the enhanced fluid circulation and mixing achieved with the double impeller configuration. The velocity contours reveal a more uniform distribution of flow velocities throughout the bioreactor, indicating improved fluid movement and mixing efficiency. The presence of two impellers working in tandem generates stronger fluid agitation and promotes better fluid dispersion compared to a single impeller setup. Furthermore, the streamlines in the dual impeller configuration exhibit a more complex and interconnected pattern compared to the single impeller case. This suggests a higher degree of fluid mixing and transport within the bioreactor, leading to enhanced mass transfer and nutrient distribution. It is worth noting that, the similarities in agitation behavior across the three speed cases in both configurations suggest that the vent-based impellers maintain their effectiveness across various operating conditions. The velocity of flow across the three vent openings is found to be greater at higher speeds, specifically from 100 to 150 RPM.

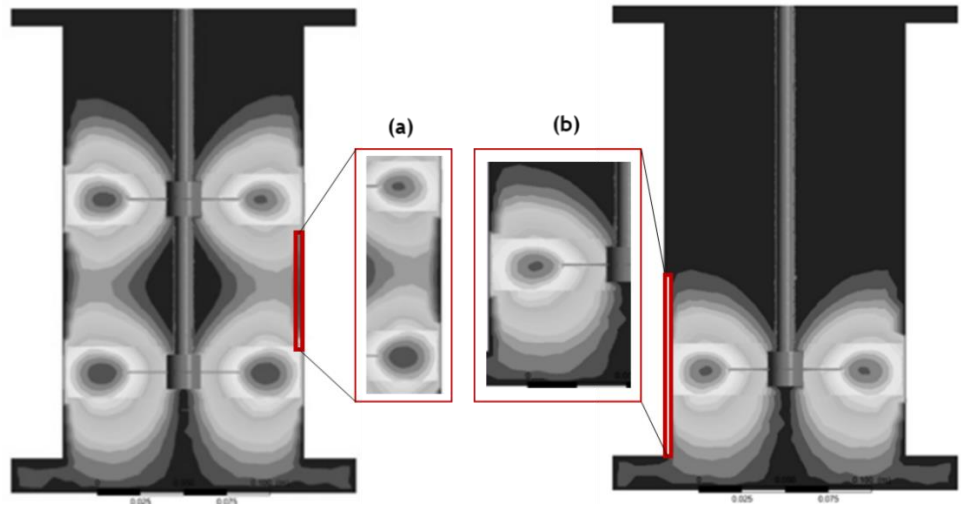

**Fig. 17.** Evidence of significant reduction and elimination of stagnant zone in (a) double impeller (b) single impeller configurations, respectively.

### 4.2 Pressure distribution in both single and double impeller bioreactors

As seen in Fig. 9, in the single impeller, the uniform and relatively high pressure within the vessel, except around the impeller region where the pressure drops, indicate efficient mixing and circulation of the fluid. The pressure drops around the impellers confirms the presence of strong fluid agitation and turbulence, which are known to enhance mixing efficiency and mass transfer rates. These results corroborate established facts from literature that emphasize the importance of optimal hydrodynamics for improved bioreactor performance (Yao *et al* 2022). Comparing the pressure distribution in the single impeller configuration to the double (Fig. 10), the irregular pressure distribution in the double impeller agitator, particularly at lower speeds, suggests the presence of complex flow patterns due to the interaction between the two impellers. The minimal pressure drops around the top impeller indicates that the base impeller is primarily responsible for the upstream work and fluid circulation, highlighting the importance of impeller positioning and interaction in multi-impeller systems. The significant pressure drops around the impeller region in both single and double impeller configurations indicate that the new impeller design has the potential to enhanced mass transfer and nutrient availability near the cells.

### 4.3 Impeller and shaft wall pressure in both single and double impeller configuration

As presented in the results (Fig. 12 and 13a-c), the observed pressure patterns on the impeller blades and shaft walls for the single impeller reveal distinct behaviors at different speeds. The higher pressure at the fluid cutting vertices of the impeller blades indicate intense fluid agitation and shear forces, which are essential for effective mixing and mass transfer. The lowest pressure observed around the rim connecting the impeller to the shaft suggests that this region experiences lower fluid forces and may result in reduced shear stress on the cells. As the speed increases in the single impeller bioreactor, the pressure on the impeller blades reduces, indicating a decrease in the intensity of fluid agitation. This observation is expected, as higher rotational speeds lead to increased fluid shear and mixing efficiency, thereby reducing the pressure on the impeller blades. The pressure drop at the impeller-shaft connection becomes more pronounced at higher speeds, indicating a potential area of concern for shear stress generation. These results emphasize the need for careful consideration of impeller-shaft design to minimize shear stress and optimize bioreactor performance.

### 4.4 Wall shear stress on impellers and shafts in both single and dual agitation

The implications of the wall shear stress results are crucial for understanding the mechanical forces experienced by the impellers and shafts and their potential impact on cell viability and product quality. Excessive shear stress can result in cell damage, decreased viability, and altered product characteristics (Ebrahimi *et al* 2019). Therefore, optimizing impeller design and operating conditions to minimize shear stress is essential for maintaining the integrity of the bioprocess. Here, in the single impeller agitation, the observed increase in wall shear stress around the impeller areas as the speed increases (see Fig. 14a-c) aligns with established facts from literature on previous designs. Higher rotational speeds lead to increased turbulence and shear forces, resulting in elevated WSS values (Salman *et al* 2021). The concentration of maximum WSS around the edges and tips of the impeller blades, which are in continuous contact with the fluid, indicates the regions of highest shear stress. However, the presence of vents and the design of the impeller blade width help distribute the flow and reduce the severity of shear stress on the impeller. The less significant effect of WSS on the entire shaft wall, which experiences high frictional tangential forces, was due to the material selection of steel. Comparing the WSS on the shafts and impellers in both the single and double impeller configurations (see Fig. 15a-c), the WSS on both the impeller and shaft increases with speed. However, the analysis reveals a noticeable difference in the WSS behavior between the top and base impellers in the double configuration. The wall shear stress is heaviest on the top impellers, particularly at higher speeds, due to the workload on the top impeller as the base impeller displaces the fluid upstream against its walls. This differential WSS distribution between the impellers highlights the interaction and dynamic forces between the impellers in the double impeller configuration, and underscores the significance of impeller arrangement and fluid dynamics in multi-impeller systems, to minimize shear stress-related issues.

### 4.5 Analysis of eddy viscosity, turbulence eddy dissipation, and turbulence kinetic energy

In the case of the single impeller configuration, it was observed that the eddy viscosity decreased with increasing speed, reaching similar ranges at speeds of 50 and 100 RPM (Fig.16a). This suggests that as the impeller rotates at higher speeds, it imparts greater momentum to the surrounding fluid, resulting in increased turbulence and mixing. This increased turbulence disrupts the formation of eddies, leading to a decrease in eddy viscosity. Similarly, in the double impeller configuration (Fig. 16a; right), the slight difference in eddy viscosity between the top and base impellers can be attributed to the complex interaction between the two impellers and the fluid flow. The closer relationship between speed variants in the double impeller configuration suggests a more synchronized flow pattern, potentially contributing to improved flow characteristics and turbulence control. The analysis of turbulence eddy dissipation and turbulence kinetic energy revealed similar behavior in both the single and double impeller configurations at the different speeds (see Fig. 16b and c). The dissipation pattern followed an upward-downward trend, contrasting with the downward-upward pattern observed in eddy viscosity. Both impeller configurations displayed a relatively uniform pattern of eddy dissipation, although the double impeller configuration exhibited higher dissipation at the top impeller with increasing speed. Furthermore, the turbulence kinetic energy patterns differed between the top and base impellers. The base agitation displayed increasing convex patterns at speeds of 50, 100, and 150 RPM, while the top impeller exhibited a more uniform straight pattern at speeds of 100 and 150 RPM, but formed a convex shape similar to the base impeller at 50 RPM. These findings suggest that the vent-based impeller design has the potential to enhance turbulence kinetic energy and promote more effective mixing in the bioreactor especially at higher speeds.

In addition, there are considerable variations in both maximal and axial velocities when comparing the vent-based impeller from the current study with the Segment Rushton (SR) and Segment-Segment (SS) impeller configurations in literature. Maximum velocities for SR and SS are often lower than those of the present vent-based impeller arrangement. For example, the maximum velocities and the axial velocities at 150 RPM for SS and SR were 0.38 m/s and 1.09 m/s, and 0.38 m/s and 0.3 m/s , respectively in Gelves *et al* (2014) and Ebrahimi *et al* (2019), whereas our design yielded 2.12 m/s and 0.52 m/s, which is a far higher value. This suggests that higher flow rates and axial flow can be attained by the vent-based impeller, which is advantageous for the reactor's vertical mixing and circulation.

### 5.0 Conclusion

This study critically evaluated the performance of a novel stirred tank bioreactor impeller that consist of vents on the four blades of the impeller, with the goal of achieving homogeneity, mitigated stagnant zone, and reduced wall shear stress. The impeller performance was evaluated in two different 3D models of lab-scale bioreactors at three different speeds: 50, 100 and 150RPM. Fluid mixing in both bioreactors was achieved through single impeller and double impeller agitations, while computational fluid dynamics was employed to simulate the hydrodynamic behavior, and computational variables within the bioreactors such as the flow velocity, velocity vector, streamlines, shaft and impeller wall shear stress and wall pressure, were obtained and compared. The velocity contours and streamlines analysis revealed that the presence of the vents in both the single and double impeller agitations enhanced flow patterns, resulting in improved homogeneity within the bioreactor. For the three speeds, the single impeller configuration yielded desirable axial flow with 0% stagnant zone, while the double impeller configuration yielded evenly distributed upward radial flow with significant reduction in stagnant zones. The analysis for eddy viscosity, turbulence eddy dissipation, and turbulence kinetic energy, suggest that the flow distribution mechanism of the vents, especially in the double impeller configuration, promotes mass, energy and particle dissipation, which are essential for cell culture. Moreover, wall shear stress on the shaft and impeller blades was significantly low, because the design of the vents parallel to the blades act as a form of flow control mechanism that helps in reducing high flow stress on the blade surfaces, exposing only a small area of the blade to flow yet achieving desirable mixing through the auxiliary agitations. This mechanism also helps in minimizing the potential for excessive turbulence or localized flow disturbances. The computational evaluation indicates that the vent-based impeller design has potential for improving turbulence control, reducing wall shear stress, enhancing mixing performance with mitigated stagnant zones, and achieving desired flow characteristics in bioreactors. Findings from this study provide valuable insights for future design and optimization of bioreactor systems, where achieving homogeneity is crucial for various applications such as cell culture, fermentation, and chemical reactions.

**Credit authorship contribution statement**

**Ayodele Oyejide:** conceptualization, study design, methodology, simulations, result post-processing and writing of initial manuscript. **Chidera Okeke**: CAD modeling, simulations, result post-processing and writing of initial manuscript. **Jesuloluwa Zacchaeus** and **Ebenezer Ige**: equally contributed to result analysis, data presentation and writing of the final manuscript. All authors read and approved the final manuscript.

**Funding**
None

**Declaration of Competing Interest**

The authors declare that they have no known competing financial interests or personal relationships that could have appeared to influence the work reported in this paper.